\documentclass[twocolumn,floatfix,preprintnumbers,nofootinbib,superscriptaddress]{revtex4}

\usepackage{ulem}
\usepackage{bm}
\usepackage{times}
\usepackage{amssymb,amsbsy,amsmath,amsfonts}
\usepackage{graphicx}
\usepackage{float}
\usepackage{color}
\usepackage{morefloats}
\usepackage{rotating}
\usepackage{srcltx}
\usepackage{slashed}
\usepackage{subfigure}
\usepackage{multirow}
\usepackage{verbatim}
\usepackage{hyperref}
\usepackage{tabularx}

\usepackage[outdir=./]{epstopdf}

\usepackage{graphicx}
\usepackage{amsmath}
\usepackage{amsfonts}
  \usepackage{   }
\usepackage{multirow}
\usepackage{mathrsfs}
\usepackage{gensymb}

\usepackage{booktabs}  
\usepackage{threeparttable}

\usepackage{subfigure}

\newcommand\be{\begin{eqnarray}}
\newcommand\ee{\end{eqnarray}}

\begin{document}

\title{Role of $\Sigma(1660)$ in the $K^- p \to\pi^0\pi^0\Sigma^0$ reaction}

\author{Xing-Yi Ji}
\affiliation{School of Physics, Zhengzhou University, Zhengzhou 450001, China}
\affiliation{State Key Laboratory of Heavy Ion Science and Technology, Institute of Modern Physics, Chinese Academy of Sciences, Lanzhou 730000, China}

\author{Si-Wei Liu}
\affiliation{State Key Laboratory of Heavy Ion Science and Technology, Institute of Modern Physics, Chinese Academy of Sciences, Lanzhou 730000, China} 
\affiliation{School of Nuclear Sciences and Technology, University of Chinese Academy of Sciences, Beijing 101408, China}

\author{Wen-Tao Lyu}
\affiliation{School of Physics, Zhengzhou University, Zhengzhou 450001, China}

\author{De-Min Li}~\email{lidm@zzu.edu.cn}
\affiliation{School of Physics, Zhengzhou University, Zhengzhou 450001, China}

\author{En Wang}~\email{wangen@zzu.edu.cn}
\affiliation{School of Physics, Zhengzhou University, Zhengzhou 450001, China}

\author{Ju-Jun Xie}~\email{xiejujun@impcas.ac.cn}
\affiliation{State Key Laboratory of Heavy Ion Science and Technology, Institute of Modern Physics, Chinese Academy of Sciences, Lanzhou 730000, China} 
\affiliation{School of Nuclear Sciences and Technology, University of Chinese Academy of Sciences, Beijing 101408, China}
\affiliation{Southern Center for Nuclear-Science Theory (SCNT), Institute of Modern Physics, Chinese Academy of Sciences, Huizhou 516000, China}

\begin{abstract}

The processes of $K^-p \to \pi^0 \pi^0 \Sigma^0$ and $K^- p \to \pi^0 \Lambda(1405)$ are studied within the effective Lagrangian approach. In addition to the “background” contribution from the $u$-channel nucleon pole term, contribution from the $\Sigma(1660)$ resonance with spin-parity $J^P=1/2^+$ is also considered. For the $K^-p \to \pi^0 \pi^0 \Sigma^0$ reaction, we perform a calculation for the total and differential cross sections by considering the contribution from the $\Sigma(1660)$ intermediate resonance decaying into $\pi^0 \Lambda(1405)$ with $\Lambda(1405)$ decaying into $\pi^0 \Sigma^0$. With our model parameters, the available experimental data on both the $K^-p \to \pi^0 \pi^0 \Sigma^0$ and $K^- p \to \pi^0 \Lambda(1405)$ reactions can be fairly well reproduced. It is shown that we really need the contribution from the $\Sigma(1660)$ resonance, and that these experimental measurements could be used to determine some properties of the $\Sigma(1660)$ resonance.

\end{abstract}

\maketitle


\section{Introduction} \label{section:introduction}

In hadron physics, the study of hyperon resonances has attracted much attention since it can help to reveal the internal structure of hadrons and deepen the understanding of the non-perturbative properties of Quantum Chromodynamics in the low energy region~\cite{Wang:2024jyk}. In the strangeness $S=-1$ sector, many theoretical researches have been conducted by using various approaches, including the chiral unitary method~\cite{Magas:2005vu,Jido:2003cb,Oset:2001cn,Oset:1997it,Wang:2016dtb,Li:2025exm,Li:2024rqb}, quark models~\cite{Xie:2014zga,Helminen:2000jb,Jaffe:2003sg,Zhong:2013oqa}, amplitude analyses~\cite{Fernandez-Ramirez:2015tfa,Fernandez-Ramirez:2015fbq,Kamano:2015hxa,Kamano:2014zba,Zhang:2013sva,Zhang:2013cua}, lattice QCD~\cite{Hall:2016kou,Hall:2014uca,Engel:2013ig,Edwards:2012fx,Engel:2010my}, and the effective Lagrangian framework~\cite{Xie:2013mua,Xie:2013msa,Xie:2013wfa,Wei:2021qnc,Lyu:2023oqn,Wang:2014jxb}. These theoretical frameworks have successfully described many established $\Lambda$ and $\Sigma$ excited states and have predicted dynamically generated resonances arising from meson–baryon interactions. On the experimental side, antikaon–nucleon scattering has been proven to be a particularly valuable probe for investigating these hyperons, owing to its rich coupled-channel structure, such as $\pi \Lambda$, $\pi \Sigma$, and $\eta \Lambda$ channels~\cite{Khemchandani:2018amu,Guo:2012vv,Feijoo:2018den,Shi:2014vha,Gao:2012zh}.

The $\Sigma(1660)$ state, listed in the Particle Data Group (PDG) with quantum numbers $J^{P}=1/2^{+}$ and a three-star rating~\cite{ParticleDataGroup:2024cfk}, has attracted significant interest in hadron physics. In Ref.~\cite{Gao:2012zh}, the differential cross sections and $\Lambda$ polarization data for the reactions $K^{-}n \to \pi^{-}\Lambda$ and $K^{-}p \to \pi^{0}\Lambda$ were analyzed within the effective Lagrangian approach over the center-of-mass energy range of $1550\sim 1676\ \mathrm{MeV}$. It was concluded that the results clearly support the existence of $\Sigma(1660)$ in these reactions. In Ref.~\cite{Shi:2023xfz}, a convolutional neural network was employed to study the $\Sigma$ hyperons using experimental data from the $K^-p\to\pi^0\Lambda$ reaction, and the results were found to support the existence of the three-star $\Sigma(1660)$  . Complementary support for the $\Sigma(1660)$ resonance in the $K_L p \to \pi^+ \Sigma^0$ reaction was also provided in a recent work in Ref.~\cite{Guo:2025mha}. However, in Ref.~\cite{Zhong:2013oqa}, the reactions $K^-p\to\Sigma^0\pi^0$, $\Lambda\pi^0$ , and $\bar{K}^0n$ were studied at low energies by using the chiral quark-model approach, and no clear evidence for the $\Sigma(1660)$ was found in the $K^-p\to\Lambda\pi^0$ reaction. Subsequently, $\bar{K}N$ scattering was analysed within the unitary multichannel model~\cite{Fernandez-Ramirez:2015tfa}, and no evidence for the three-star state $\Sigma(1660)$ was found. It is worth mentioning that, in Ref.~\cite{MartinezTorres:2007sr}, the Faddeev equations were solved within the coupled channel approach, and a peak at total energy $\sqrt{s} = 1656$~MeV was found in the $\pi\pi\Sigma$ system, which was associated with $\Sigma(1660)$. Therefore, three-body decay channels, such as $\pi\bar{K}N$ and $\pi\pi\Sigma$, might be important for establishing the existence and properties of $\Sigma(1660)$ resonance.

The total and differential cross sections of the process $K^{-}p \to \pi^{0}\pi^{0}\Sigma^{0}$ reaction were measured with high statistics by the Crystal Ball Collaboration at incident kaon momenta from $514$ to $750\ \mathrm{MeV}/c$~\cite{CrystallBall:2004ovf}. It is shown that this reaction is dominated by the $\pi^0 \Lambda(1405)$ intermediate state. Thus, these experimental measurements offer an opportunity to study the existence and properties of the $\Sigma(1660)$ through the three-body decay mode of $\Sigma(1660) \to \pi^0 \Lambda(1405) \to \pi^0 \pi^0 \Sigma^0$.

In this work, we investigate the role of $\Sigma(1660)$ resonance in the $K^{-}p \to \pi^{0}\pi^{0}\Sigma^{0}$ and $K^- p \to \pi^0 \Lambda(1405)$ reactions near threshold within the effective Lagrangian approach. In our calculation, the $s$-channel is considered to include the mechanism $\Sigma^0(1660) \to \pi^{0}\Lambda(1405) \to \pi^0\pi^{0}\Sigma^{0}$, where we have introduced model parameters of coupling constants $g_{\Sigma(1660)\bar{K}N}$ and $g_{\Sigma(1660)\pi\Lambda(1405)}$. Their values will be determined by fitting them to the total and differential cross sections of $K^- p \to \pi^0\pi^0\Sigma^0$ reaction. For the $u$-channel, the non-resonant background contribution from nucleon pole exchange is considered. In addition, the total cross section for the $K^-p\to\pi^0\Lambda(1405)$ process is also calculated to provide information on the complementary channel in the $\Sigma(1660)$ resonance region. 


This article is organized as follows. In Section~\ref{sec:formalism}, the theoretical formalism for calculating the cross sections of the processes $K^-p \to \pi^0\pi^0\Sigma^0$ and $K^-p \to \pi^0\Lambda(1405)$ is presented. Numerical results and discussions are shown in Section~\ref{sec:results}. Finally, a short summary is given in the last section.

\section{Theoretical formalism} \label{sec:formalism}

\subsection{The $K^-p \to \pi^0\pi^0\Sigma^0$ reaction}

\begin{figure}[htbp]
\centering
\subfigure[]{\includegraphics[scale=0.32]{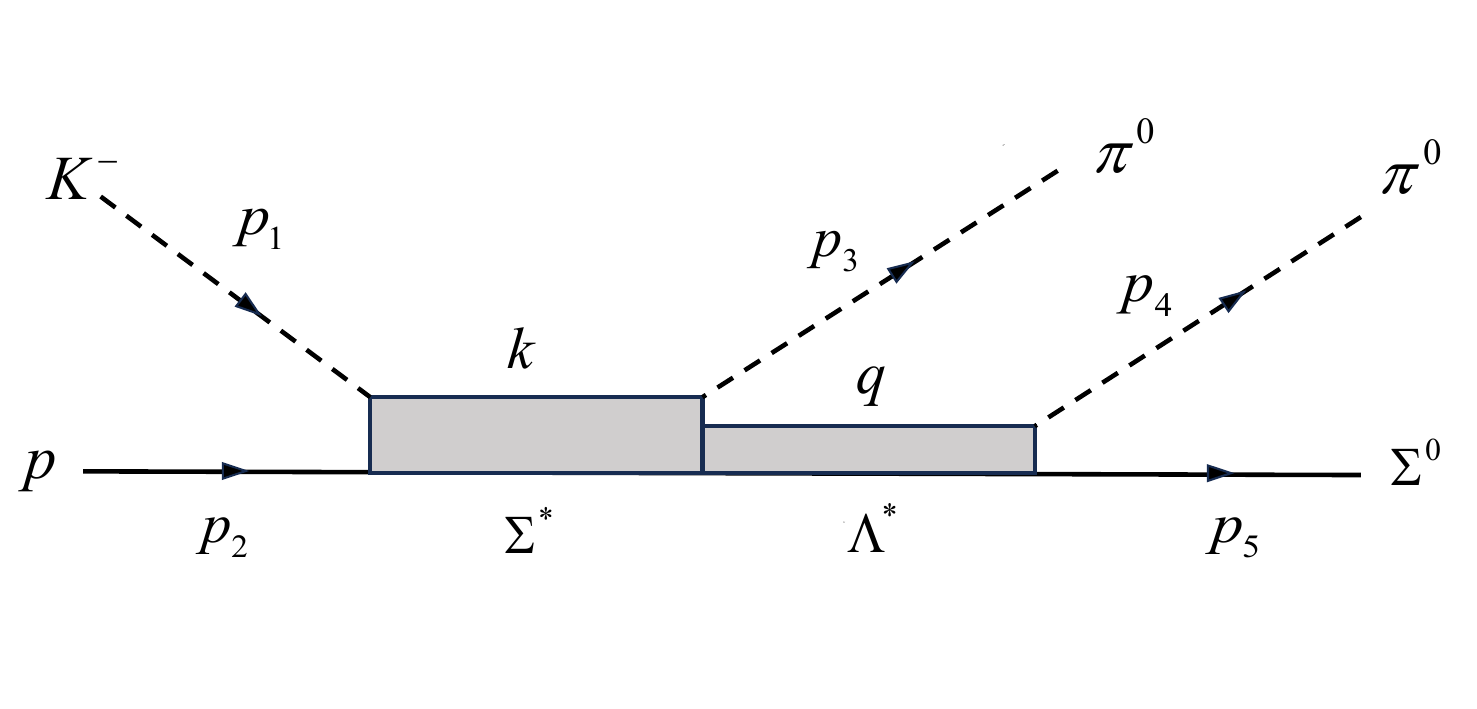}
\label{fig:Sigma}}
\subfigure[]{\includegraphics[scale=0.32]{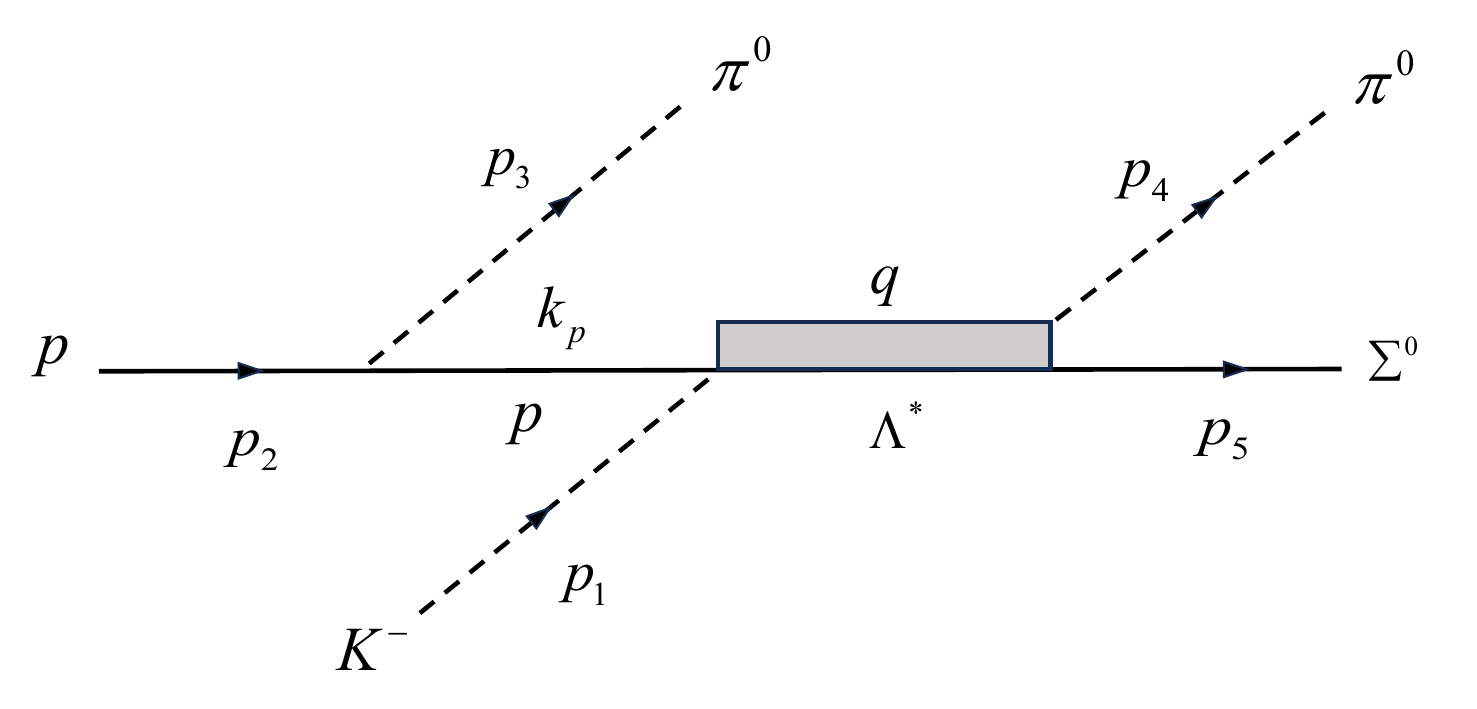}
\label{fig:HX}}
\caption{Scattering mechanisms of the $K^-p\rightarrow\pi^0\pi^0\Sigma^0$ reaction. It consists of $s$-channel $\Sigma(1660)$ resonance (a), and $u$-channel nucleon pole term (b). We also show the definition of the kinematical variables ($p_1$, $p_2$, $p_3$, $p_4$, $p_5$, $k$, $k_p$, $q$) used in the calculation.}
\label{fig:feynman}
\end{figure}

In this section, we show the theoretical formalism for studying the $K^-p \to \pi^0\pi^0\Sigma^0$ within the framework of effective Lagrangian method~\cite{Zhou:2019eaf,Lyu:2023oqn,Dai:2025hvo,Wang:2025cex}, where these combined contributions from both the $s$-channel and $u$-channel are considered, as illustrated in Fig.~\ref{fig:feynman}. For the $s$-channel, the mechanism of $K^-p\to\Sigma(1660)\to\pi^0\Lambda(1405)\to\pi^0\pi^0\Sigma^0$ is analyzed, while for the $u$-channel, we include the contribution of proton pole term. 

To evaluate the contribution of $s$-channel shown in Fig.~\ref{fig:feynman} (a), the effective Lagrangian densities are introduced for these interaction vertices~\cite{Gao:2012zh,Xie:2013wfa,Kim:2024mqx},
 \be 
    & \mathcal{L}_{\bar{K} N\Sigma^*}&=-i{g}_{\bar{K} N\Sigma^*}\bar{N}\gamma_5 \tau \cdot \vec{\Sigma} K + \text{H.c.} , \label{eq:L_KNSigma*}\\
    &\mathcal{L}_{\Sigma^*\pi\Lambda^*}&=ig_{\Sigma^*\pi\Lambda^*}
        \bar{\Lambda}^*\vec\pi \cdot \vec\Sigma^* +\text{H.c.},\label{eq:L_Sigma*piLambda}\\    
    & \mathcal{L}_{\Lambda^*\pi\Sigma}&=-ig_{\Lambda^*\pi\Sigma}
     \bar{\Lambda}^*\vec{\pi}\cdot\vec{\Sigma}+\text{H.c.}\label{eq:L_LambdapiSigma},
\ee
where the $\Sigma^*$ denotes the $\Sigma(1660)$ resonance with spin-parity $J^P = 1/2^+$ and $\Lambda^*$ denotes the $\Lambda(1405)$ resonance. The operator $\tau$ represents the Pauli matrix (see Ref.~\cite{Gao:2012zh}). $g_{\bar{K} N\Sigma^*}$, $g_{\Sigma^*\pi\Lambda^*}$ , and $g_{\Lambda^*\pi\Sigma}$ are the couplings for the vertices $\bar{K} N\Sigma^*$, $\Sigma^*\pi\Lambda^*$ , and $\Lambda^*\pi\Sigma$, respectively. In this work, the product $g_{\Sigma^*\bar{K} N}g_{\Sigma^*\pi\Lambda^*}$ is determined by fitting to the experimental data.

On the other hand, the non-resonant background contribution from the $u$-channel nucleon pole is also considered, as depicted in Fig.~\ref{fig:feynman}(b). The contribution of this part is calculated by following the same approach. It should be noted that the effective Lagrangian density for the $\Lambda^*\pi\Sigma$ vertex has been written in Eq.~(\ref{eq:L_LambdapiSigma}), while those for the other vertices are given by~\cite{Gao:2012zh,Xie:2013wfa,Kim:2024mqx}
\be
    & \mathcal{L}_{\pi NN}&=-i{g}_{\pi NN}
        \bar{N}\gamma_5\vec{\pi}\cdot\tau N+\text{H.c.}, \label{eq:piNN}\\
    & \mathcal{L}_{\bar{K}N\Lambda^*}&=ig_{\bar{K} N\Lambda^*}
        \bar N\bar{K}{\Lambda}^*+\text{H.c.},\label{eq:KNLambda}
\ee 
where $g_{\pi NN}$ and $g_{\bar{K}N\Lambda^*}$ are the coupling constants for the vertices $\pi NN$ and $\bar{K}N\Lambda^*$, respectively. In our calculation, the values $g_{\pi NN}=13.45$, $g_{\Lambda^*\pi\Sigma}=0.9$ and $g_{\Lambda^*\bar{K}N}=1.51$ are taken from Ref.~\cite{Xie:2013wfa}, where these values were obtained by fitting them to the experimental data
on the total cross sections of the $K^- p \to \pi^0\Sigma^0$ reaction.

With the effective-interaction Lagrangian densities given above, we can easily construct the invariant scattering amplitudes:
\begin{eqnarray}
\mathcal{M}_s &=&  -g_{\Lambda^*\pi\Sigma}g_{\Sigma^* \pi \Lambda^*}g_{\bar{K} N\Sigma^*} 
    F_{\Sigma^*}(k^2)F_{\Lambda^*}(q^2) \nonumber \\
       & & \bar{u}(p_{5})G_{\Lambda^*}(q) G_{\Sigma^*}(k) \gamma_5u(p_{2}), \label{eq:M_s_channel} \\
\mathcal{M}_u &=& -g_{\Lambda^*\pi\Sigma} g_{\bar{K}N\Lambda^*}g_{\pi NN} F_{p}(k_p^2)F_{\Lambda^*}(q^2) \nonumber \\
       & & \bar{u}(p_{5})G_{\Lambda^*}(q)G_p(k_p)\gamma_5u(p_{2}), \label{eq:M_u_channel}
\end{eqnarray}
where $G_R(q)$ [$R \equiv \Sigma(1660)$, $\Lambda(1405)$ or proton pole] is the propagator for spin-$\frac{1}{2}$ particle~\cite{Lu:2014rla,Dai:2025hvo},
\begin{equation}\label{eq:G}
    \begin{aligned}
        G_{R}(q)&=\frac{i(\slashed q+m_R)}{q^2-m_R^2+im_R\Gamma_R},
    \end{aligned}
\end{equation}
with $m_R$ and $\Gamma_R$ the mass and total decay width of the intermediate resonance, respectively. We take $m_{\Sigma(1660)} = 1660$~MeV, $\Gamma_{\Sigma(1660)} = 200$~MeV, $m_{\Lambda(1405)} = 1405$~MeV, and $\Gamma_{\Lambda(1405)} = 50.5$~MeV~\cite{ParticleDataGroup:2024cfk}. Moreover, it is necessary to introduce the form factors to account for the off-shell behavior of the intermediate particles~\cite{Gao:2010ve,Liu:1995st,Xie:2007qt,Xie:2015zga,Feuster:1998cj}. In this work, the following form factor is adopted~\cite{Xie:2010yk,Gao:2010ve,Wang:2016fhj,Wang:2017hug},
\begin{equation}\label{eq:F}
    \begin{aligned}
        F_R(q^2)=\frac{\Lambda_R^{4}}{\Lambda_R^{4}+\left(q^{2}-m_R^2\right)^{2}},
    \end{aligned}
\end{equation}
where $\Lambda_R$, $m_R$, and $q$ are the cutoff parameter, mass, and four-momentum of the intermediate baryon, respectively. In this work, the cutoff parameters $\Lambda_{\Sigma(1660)}$, $\Lambda_{\Lambda(1405)}$, and $\Lambda_p$ are constrained between 800 to $1200$~MeV.

Using the formalism mentioned above, the differential cross section for $K^-p \to \pi^0\Lambda(1405) \to \pi^0\pi^0\Sigma^0$ can be constructed as follows~\cite{Xie:2015zga},
\begin{equation}\label{eq:three-body_differential_cross_sections}
    \begin{aligned}
        d\sigma &= \frac{|\vec{p}_{3}^{~*}| |\vec{p}_{5}|}{2^{12} \pi^5s |\vec{p}_1| }dm_{12} d\Omega_1 d\Omega_2^*\times  \sum_{\text{spins}}|\mathcal{M_{\text{total}}}|^2,
    \end{aligned}
\end{equation}
where we have considered the factor of $1/2$ accounting for the identity of the two final-state $\pi^0$, and also the factor $1/2$ for averaging over the spins of the initial state. In the calculation, the momentum exchange between the two final $\pi^0$ mesons has also been taken into account.

\begin{figure}[htbp]
\centering
\includegraphics[scale=0.42]{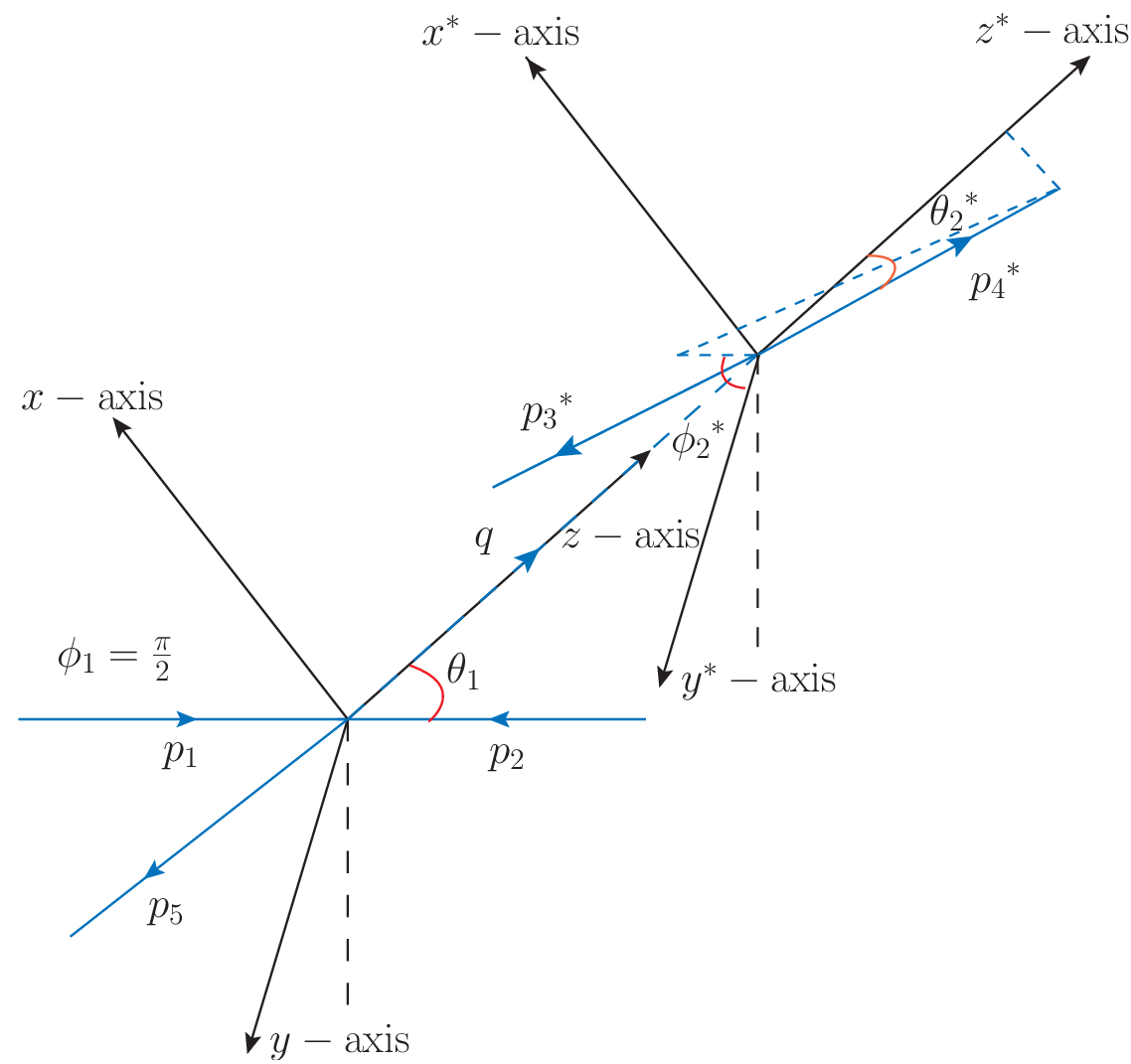}
\caption{Definition of the three-body phase space in the $K^-p\rightarrow\pi^0\pi^0\Sigma^0$ reaction.} \label{fig:angle}
\end{figure}

The invariant mass square of the $K^- p$ system can be obtained by
\begin{equation}
s=\left(\sqrt{m_{K^-}^2+|\vec{p}_{K^-}|^2}+m_{p}\right)^2-|\vec{p}_{K^-}|^2,
\end{equation}
where $|\vec{p}_{K^-}|$ is the kaon momentum in the laboratory frame. $|{\vec{p}_{3}}|$ and $ |\vec{p}_5^{~*}|$ in Eq.~(\ref{eq:three-body_differential_cross_sections}) stand for the momenta of $\pi^0$ in the rest frame of the $\Sigma(1660)$ and $\Lambda(1405)$ resonances, respectively, and they are given by
\be 
|\vec{p}_3^{~*}| &=& \frac{\lambda^{ \frac{1}{2}}(m_{12 }^2,m_{\pi^0}^2,m_{\pi^0}^2)}{2m_{12 }},\\
 |{\vec{p}_{5}}| &=& \frac{\lambda^{ \frac{1}{2}}(s,m_{\Sigma^0}^2,m_{12 }^2)}{2\sqrt{s}},
\ee
with the \text{Källén} function $\lambda(x,y,z) = x^{2} + y^{2} + z^{2} - 2xy - 2xz - 2yz$. In addition, the definition of the solid angles $d\Omega_1=d\cos\theta_1 d\phi_1$ and $d\Omega_2^*=d\cos\theta_2^* d\phi_2^*$ in Eq. (\ref{eq:three-body_differential_cross_sections}) is shown as in Fig.~\ref{fig:angle}.

Finally, the squared modulus of the total scattering amplitude is given by
\begin{equation}
    \begin{aligned}
    |\mathcal{M}_{\text{total}}|^2&= |\mathcal{M}_s+\mathcal{M}_u|^2.
    \end{aligned}
\end{equation}

\subsection{The $K^-p\rightarrow\pi^0\Lambda(1405)$ reaction}

\begin{figure}[htbp]
\centering
\subfigure[]{\includegraphics[scale=0.4]{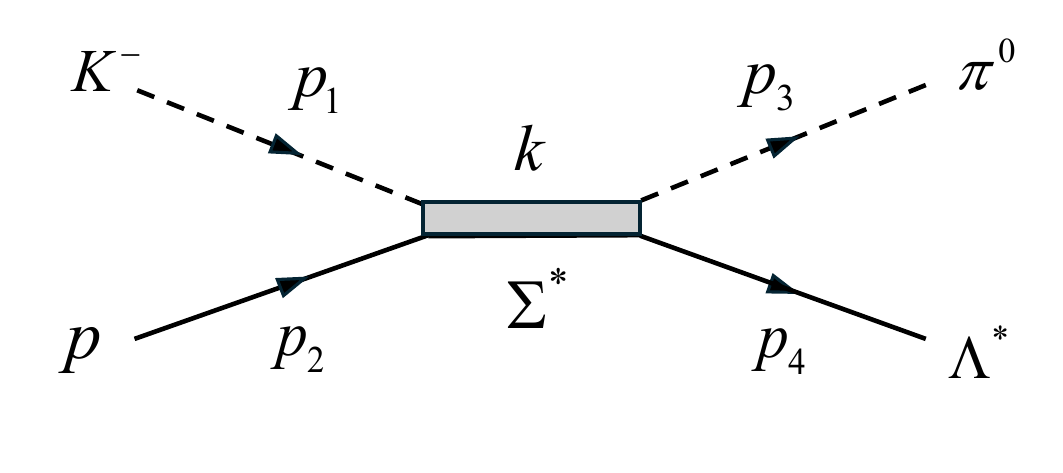}
}
\subfigure[]{\includegraphics[scale=0.4]{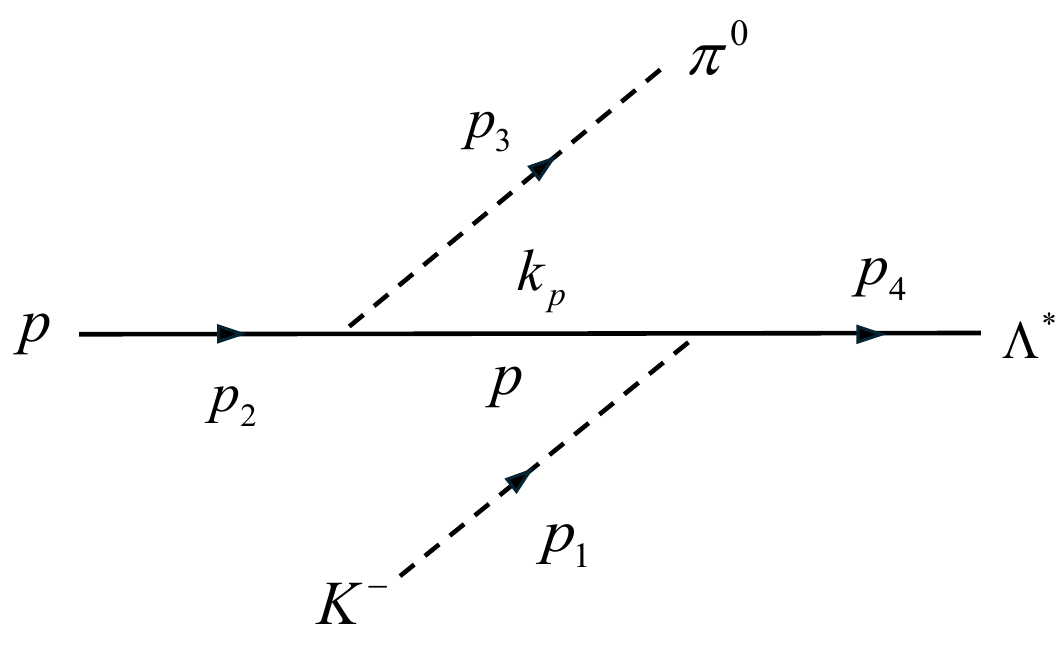}
}
\caption{Scattering mechanisms for the $K^-p \to \pi^0\Lambda(1450)$ reaction. (a): $s$-channel, (b): $u$-channel.}
\label{fig:feynman-twobody}
\end{figure}

To further constrain the model and provide complementary information, the related two-body scattering process $K^-p \to \pi^0\Lambda(1405)$ is also investigated within the same framework. The relevant reaction diagrams are shown in Fig.~\ref{fig:feynman-twobody}. Based on the effective Lagrangian densities for the relevant vertices in Eqs.~(\ref{eq:L_KNSigma*}), (\ref{eq:L_Sigma*piLambda}), (\ref{eq:piNN}), and (\ref{eq:KNLambda}), the scattering amplitudes for the $K^-p \to \pi^0\Lambda(1405)$ reaction are written as
\begin{eqnarray}
\Tilde{\mathcal{M}}_{s} &=& - g_{\Sigma^*\pi\Lambda^*}g_{\bar{K} N\Sigma^*}F_{\Sigma^*}(k^2) \nonumber \\
&& \times \bar{u}(p_4) G_{\Sigma^*} (k)\gamma_5 u(p_{2}),\\
\Tilde{\mathcal{M}}_u &=& -g_{\bar{K}N\Lambda^*}g_{\pi NN}F_{p}(k_p^2) \nonumber \\
&& \times \bar{u}(p_4)G_p(k_p)\gamma_5u(p_{2}).
\end{eqnarray}

Then, the differential cross section for the $K^-p\rightarrow\pi^0\Lambda(1405)$ reaction is given by~\cite{ParticleDataGroup:2024cfk}
\be 
   d\sigma = \frac{|\vec{p}_3|}{64\pi|\vec{p}_1|s } \sum_{\text{spins}} |\Tilde{\mathcal{M}}_s+\Tilde{\mathcal{M}}_u|^2 d{\rm cos}{\Tilde\theta},
\ee
where $\tilde{\theta}$ is the scattering angle of the outgoing $\pi^0$ meson.

\section{Numerical results and discussions}\label{sec:results}

With the formalism and ingredients given above, the total and differential cross sections versus the $K^-$ momentum $p_{K^-}$ for the $K^-p \to \pi^0\pi^0 \Sigma^0$ reaction are calculated. Because the experimental data are limited and have large uncertainties, we should try to reduce the number of free theoretical parameters. In this work, the cutoff parameters $\Lambda_{\Sigma(1660)}$, $\Lambda_{\Lambda(1405)}$, and $\Lambda_p$ are all fixed to $1000$~MeV.~\footnote{The affection of varying the cutoff parameters have also been tested, but no significantly better fits were obtained.} The masses and widths of these involved particles are taken as the nominal values quoted in the PDG~\cite{ParticleDataGroup:2024cfk}. Among these values, by considering that $\Sigma(1660)$ is a broad state and its width has not yet been established~\cite{ParticleDataGroup:2024cfk}, the center value $\Gamma_{\Sigma(1660)}=200$~MeV is adopted. Then, there is only one free parameter $g_{\Sigma^*\bar{K}N}g_{\Sigma^*\pi\Lambda^*}$ in our model, which corresponds to the combined coupling strength of the $\Sigma(1660)\bar{K} N$ and $\Sigma(1660)\pi\Lambda(1405)$ vertices. Its value is determined by a combined fit to the data for the total and differential cross sections of the $K^-p \to \pi^0\pi^0\Sigma^0$ process and the total cross section of the $K^-p \to \pi^0 \Lambda(1405)$ reaction. There are a total of 77 data points. The value of $g_{\Sigma^*\bar{K}N}g_{\Sigma^*\pi\Lambda^*}$ is determined as $1.81 \pm 0.10$, with a reasonable small $\chi^2/ \rm{d.o.f} = 1.46$.

\begin{figure}[htbp]
\centering
\includegraphics[scale=0.45]{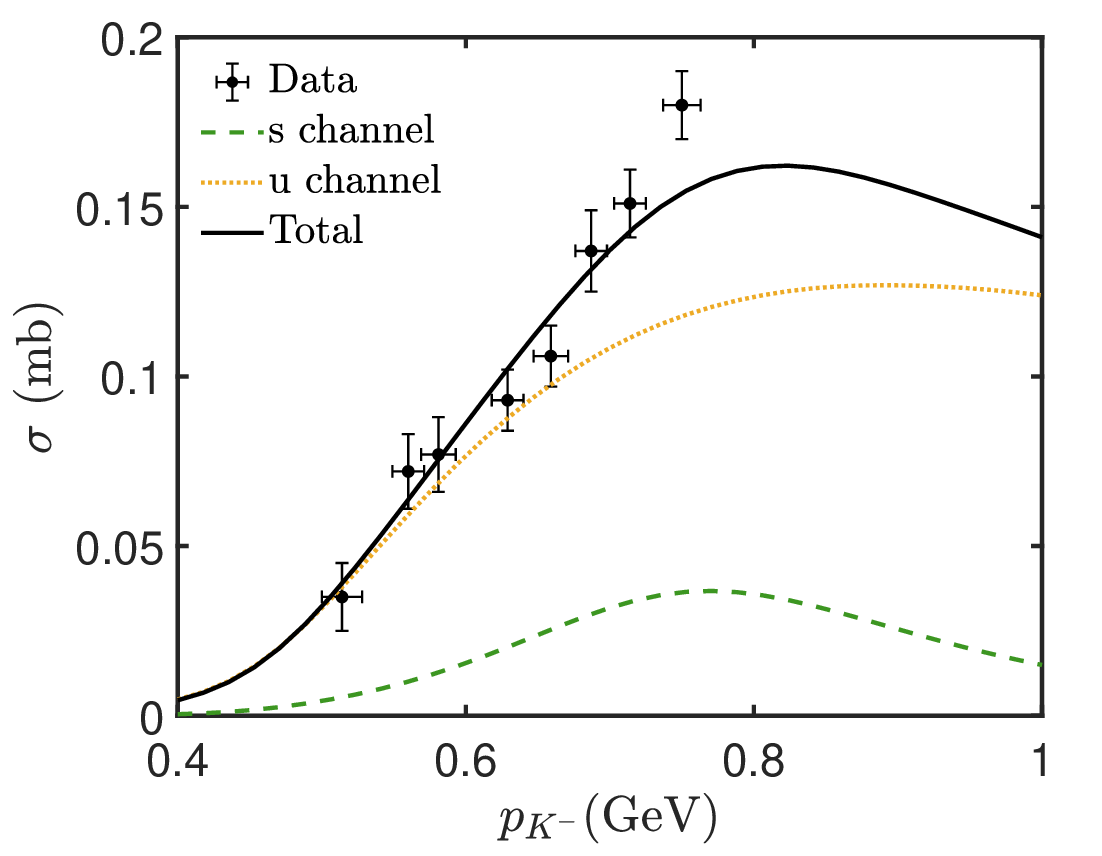}
\caption{Total cross section for the $K^-p \rightarrow \pi^0\pi^0\Sigma^0$ reaction. The experimental data are taken from Ref.~\cite{CrystallBall:2004ovf}.} \label{fig:total_cross_section_threebody.pdf}
\end{figure}

The theoretical fitted total cross sections for the $K^-p \to \pi^0\pi^0\Sigma^0$ process are shown in Fig.~\ref{fig:total_cross_section_threebody.pdf}, where the kaon momentum $p_{K^-}$ ranges from $0.4$ to $1.0$ GeV, corresponding to a center-of-mass energy $\sqrt{s}$ ranging from $1.52$ to $1.79$~GeV. The green-dashed curve represents the contribution from the $s$-channel $\Sigma(1660)$ resonance. The yellow-dotted curve stands for the contribution from the $u$-channel nucleon pole. The black-solid curve corresponds to their total contributions. One can see that the $u$-channel nucleon pole is dominant. This is in agreement with the results of Ref.~\cite{Magas:2005vu}. However, the possible broad structure around $p_{K^-}=0.8$~GeV is also observed, which is contributed by the $\Sigma(1660)$ resonance. In other words, the $\Sigma(1660)$ contributes to the high-energy region of the total cross section and plays an important role in this reaction. It is expected that more experimental data can be used to check our predictions, and to better understand the properties of the $\Sigma(1660)$ resonance.

\begin{figure*}[htbp]
\centering
\includegraphics[scale=0.6]{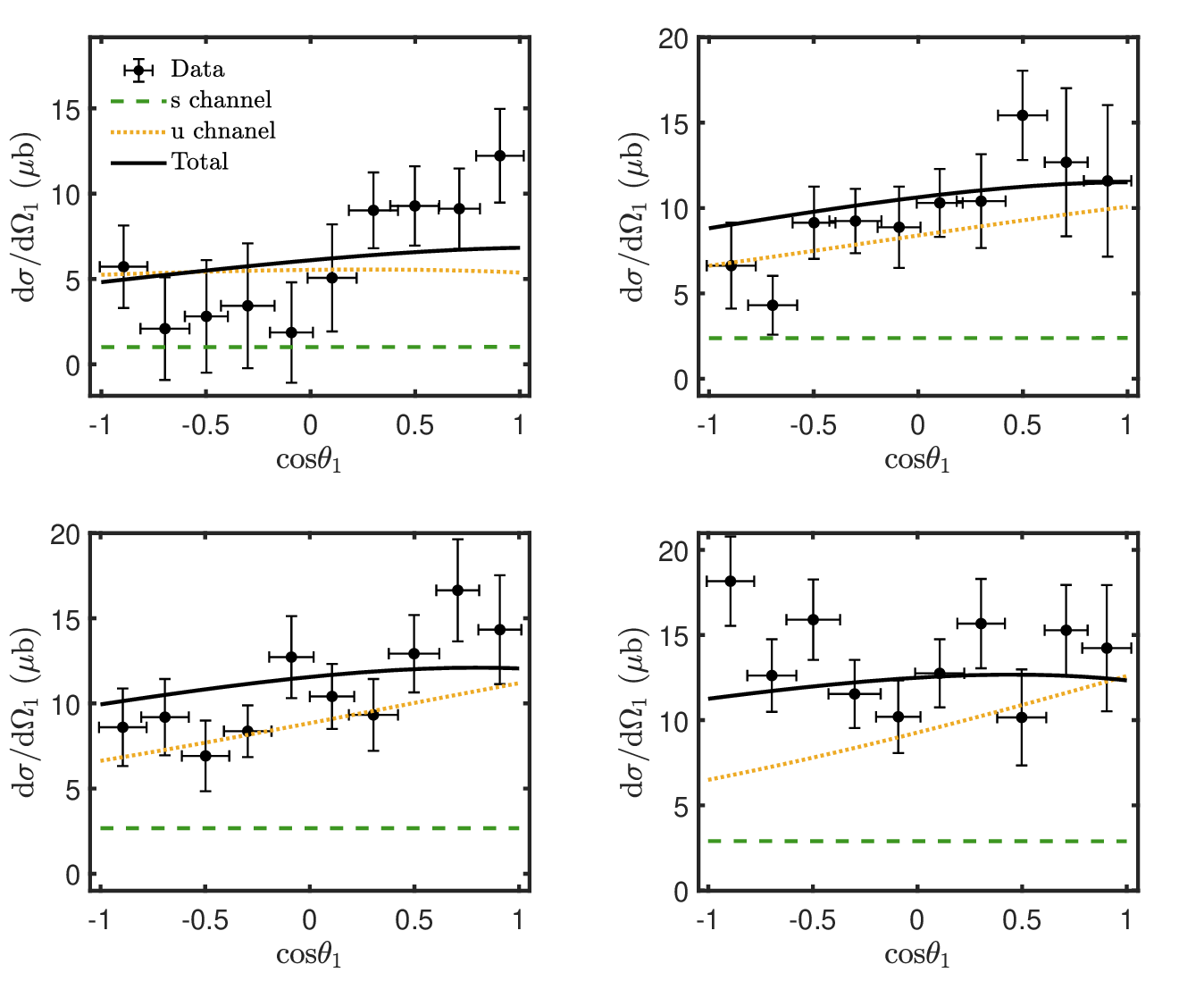}
\caption{Differential cross sections of $K^-p\rightarrow\pi^0\pi^0\Sigma^0$ reaction for the $p_{K^-}=518$~MeV~(upper left), 687~MeV~(upper right), 714~MeV~(lower left) and 750~MeV~(lower right). The experimental data are taken from Ref.~\cite{CrystallBall:2004ovf}.} \label{fig:differential_cross_section2.pdf}
\end{figure*}

The differential cross sections for the $K^{-}p \to \pi^{0}\pi^{0}\Sigma^{0}$ process are also calculated and compared with data. The angular distributions for kaon momenta $p_{K^-}=514,687,714,750$~MeV are presented in Fig. \ref{fig:differential_cross_section2.pdf}, where $\theta_1$ denotes the angle between the combined momentum direction of the two outgoing pions and the incoming $K^-$ meson direction. It can be seen that our results are in fair agreement with the experimental measurements from the Crystal Ball Collaboration~\cite{CrystallBall:2004ovf}.

For the $K^-p \to \pi^0\pi^0 \Sigma^0$ reaction, the contribution from scalar meson $f_0(500)$ should be important, since the principle decay of $f_0(500)$ is into two pions. However, the meson $f_0(500)$ is a very broad state, and the current data are limited, thus we leave the study of the meson $f_0(500)$ contribution for future studies when more experimental data are available. In fact, it was pointed out that the contribution from $f_0(500)$ meson to the $K^-p \to \pi^0 \pi^0 \Sigma^0$ is insignificant in the experimental paper~\cite{CrystallBall:2004ovf}.

\begin{figure}[htbp]
\centering
\includegraphics[scale=0.45]{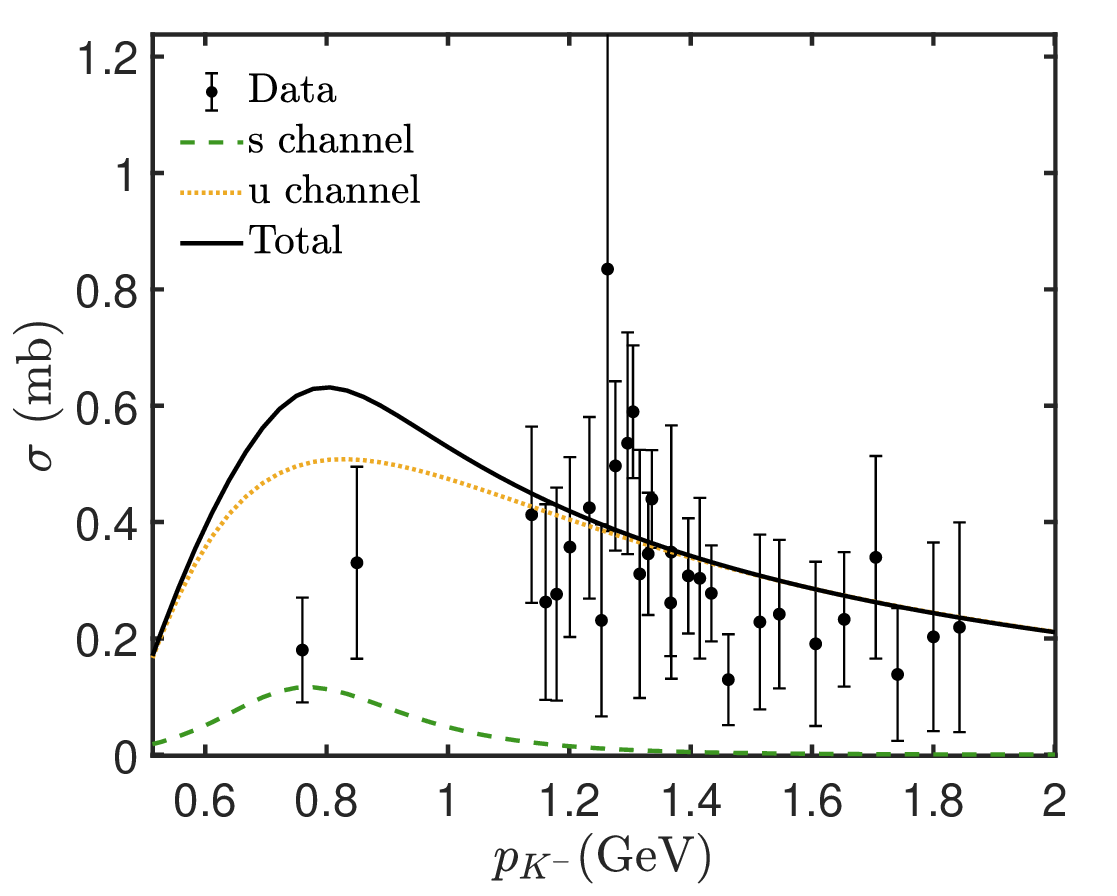}
\caption{Total cross section for the $K^-p\rightarrow\pi^0\Lambda(1405)$ reaction. The experimental data are taken from Refs.~\cite{Bastien:1961zz,Griselin:1975pa,Gordon:1969qd,Berthon:1974kr}.} \label{fig:total_cross_section_twobody.pdf}
\end{figure}

To provide information on the complementary channel, the total cross section for the $K^-p \to \pi^0\Lambda(1405)$ reaction has been analyzed, with the results presented in Fig.~\ref{fig:total_cross_section_twobody.pdf}. Here, $p_{K^-}$ ranges from $0.5$ to $2.0$ GeV corresponding to $\sqrt{s}$ from $1.56$ to $2.23$ GeV. A significant peak is observed around $p_{K^-}=0.75$~GeV. However, it is found that the available experimental data are mainly concentrated in the higher-energy region of this reaction, whereas the signal associated with the $\Sigma(1660)$ state lies in the lower-energy region. Consequently, data in the low-energy region of this reaction are crucial for confirming the properties of the $\Sigma(1660)$.

Note that, within the chiral unitary approach and by considering the two pole structures of the $\Lambda(1405)$ state~\cite{Kaiser:1995eg,Oset:1997it,Oller:2000fj,Garcia-Recio:2002yxy,Jido:2003cb,Garcia-Recio:2003ejq,Xie:2023cej,Zhuang:2024udv}, it has been shown that the $K^-p \to \pi^0 \pi^0 \Sigma^0$ reaction is particularly suited for probing the properties of the higher-energy, narrower second pole of the $\Lambda(1405)$ resonance. This is because the process is largely dominated by a mechanism in which one $\pi^0$ is emitted before the $K^- p \to \pi^0 \Sigma^0$ transition, and the $K^- p \to \pi^0 \Sigma^0$ scattering amplitude most strongly coupled to the second pole. Nevertheless, more precise experimental data are required to draw definitive conclusions.

Next, we turn to the strong couplings of $\Sigma(1660)$ to $\bar{K}N$ and $\pi\Lambda(1405)$ channels. With the effective interactions shown in Eq.~\eqref{eq:L_KNSigma*}, one can easily get the value of $g_{\Sigma^*\bar{K}N}$ using the relevant partial decay width
\begin{equation}
 \Gamma_{\Sigma(1660) \to \bar{K}N}=\frac {g^2_{\Sigma^*\bar{K}N}}{2\pi} \frac{|\vec{p}_{K^-}|}{m_{\Sigma^*}}{(E_ p-m_p)},
 \label{Eq:Gamma_SigmatoKp}
\end{equation}
where $E_p$ and $|\vec{p}_{K^-}|$ are the energy of the proton and momentum of $K^-$ in the $\Sigma(1660)$ rest frame, respectively. With the experimental data of ${\mathcal B}[\Sigma(1660) \to \bar{K}N] = 0.10 \pm 0.05$ and $m_{\Sigma^*} = 1660 \pm 20$~MeV, and $\Gamma_{\Sigma^*} = 200 \pm 100 $~MeV, we can obtain the coupling constant:
\begin{eqnarray}
g_{\Sigma^* \bar{K}N} =2.49\pm0.88,
\end{eqnarray}
where its uncertainty is calculated from the errors of mass and width of $\Sigma(1660)$ and also the branching ratio of ${\mathcal B}[\Sigma(1660) \to \bar{K}N]$.

Subsequently, with $g_{\Sigma^*\bar{K}N}g_{\Sigma^*\pi\Lambda^*} = 1.81 \pm 0.10$, one can get  
\begin{equation}
g_{\Sigma^*\pi\Lambda^*} = 0.73 \pm0.26.
\end{equation}

Then, with the obtained strong coupling constant $g_{\Sigma^*\pi\Lambda^*}$, we have evaluated the the branching fraction of $\Sigma(1660)\to \pi \Lambda(1405)$ as
 \begin{eqnarray}
 && {\mathcal B}[\Sigma(1660)\to \pi\Lambda(1405)] = \frac{\Gamma_{\Sigma(1660) \to \pi \Lambda(1405)}}{\Gamma_{\Sigma(1660)}} \nonumber \\
 && = \frac {g^2_{\Sigma^*\pi\Lambda^*}}{4\pi\Gamma_{\Sigma^*}} \frac{|\vec{p }_{\pi^0}|}{m_{\Sigma^*}}(E_{\Lambda^*} +m_{\Lambda^*}).
 \label{Eq: branching_fraction_Gamma_SigmatopiLambda}
 \end{eqnarray}
This yields $\mathcal{B}[\Sigma(1660)\to\pi^0\Lambda(1405)] = (7.2 \pm 6.3)\%$, which is consistent with the PDG value $\mathcal{B}(\Sigma(1660)\to\pi^0\Lambda(1405)) = (4.0\pm2.0)\%$ within uncertainties.

\section{Summary} \label{sec:summary}

In this work, the process $K^-p \to \pi^0\pi^0\Sigma^0$ has been analyzed within the effective Lagrangian approach. The contributions from the $s$-channel involving the $\Sigma(1660)$ and the non-resonant background from the $u$-channel nucleon pole have been considered. The total cross section and angular distributions for the $K^-p \to \pi^0 \pi^0\Sigma^0$ process have been presented. Meanwhile, the total cross section for the $K^-p\to\pi^0\Lambda(1405)$ reaction has also been studied within the same model parameters. Our results are in good agreement with the experimental data and support the conclusion that the $\Sigma(1660)$ resonance plays a key role in the $K^-p \to \pi^0\pi^0\Sigma^0$ process with its decaying into the $\pi\Lambda(1405)$ channel. Furthermore, a significant peak has been predicted in the total cross section of $K^-p\to\pi^0\Lambda(1405)$ around $p_{K^-}=0.75$~GeV, arising from the intermediate $\Sigma(1660)$.

However, it is noted that the experimental uncertainties remain large and that data in the low-energy region of the total cross section for $K^-p\to\pi^0\Lambda(1405)$ are not yet available. These key pieces of information are crucial for confirming the properties of the $\Sigma(1660)$ resonance. Therefore, precise future measurements of these processes, such
as by the KLF Collaboration at JLab using the GlueX spectrometer~\cite{KLF:2020gai} and possible Huizhou Hadron Spectrometer at HIAF~\cite{Chen:2025ppt}, are strongly encouraged, as they would be crucial for elucidating the properties of the $\Sigma(1660)$ and determining the coupling strength for $\Sigma(1660) \to \pi\Lambda(1405)$.

\begin{acknowledgments}

This work is partly supported by the National Key R\&D Program of China under Grant No. 2023YFA1606703 and No. 2024YFE0105200, and by the National Natural Science Foundation of China under Grant Nos. 12435007, 12475086, 12192263, 12361141819, and 1252200936. This work is supported by the Natural Science Foundation of Henan under Grant Nos. 232300421140 and 252300423951,  and also the Zhengzhou University Young Student Basic Research Projects (PhD students) under Grant No. ZDBJ202522.
 
\end{acknowledgments}

\normalem
\bibliographystyle{apsrev4-1.bst}
\bibliography{ref} 

\end{document}